\documentclass[aps,prb,preprint]{revtex4}
\usepackage{graphicx}
\usepackage{epsfig}

\begin{document}

\title{A unified model for the long and high jump}

\author{O. Helene}
\affiliation{Instituto de F\'{\i}sica, Universidade de S\~{a}o Paulo, 
C.P. 66318, CEP 05315-970 S\~{a}o Paulo, SP, Brasil}
\author{M. T. Yamashita}
\affiliation{Unidade Diferenciada de Itapeva, Universidade Estadual Paulista, 
CEP 18409-010 Itapeva, SP, Brasil}

\begin{abstract}
A simple model based on the maximum energy that an athlete can produce in a
small time interval is used to describe the high and long jump.
Conservation of angular momentum is used to explain why an athlete should
run horizontally to perform a vertical jump. Our results agree with world
records.
\end{abstract}

%\pacs{87.19.St.,89.20.-a}

\maketitle

\section{introduction}

A few years ago, William Harris asked how the kinetic energy acquired from
running is converted in the long and high jump
events.\cite{HaAJP97} The question was whether athletes running
horizontally can change their velocity into one that forms an angle of
45$^\circ$ with the horizontal without changing their speed. The three
negative answers\cite{answer1,answer2,answer3} were 
that athletes cannot convert their initial horizontal velocity into a
vertical one,\cite{answer2} because it is impossible to generate the
necessary power required by the task,\cite{answer3} or equivalently,
athletes cannot sustain the necessary acceleration to acquire the vertical
velocity at takeoff.\cite{answer1} 

If the answer were positive, the center of mass of a world class athlete
running at 10\,m/s and taking off at 45$^\circ$ would go a horizontal
distance of about 10.2\,m. Because the athlete's center of mass is forward
of the front edge of the runway at takeoff and behind
the point where her heels hit the ground at landing (see Sec.~III), the
actual total jump length would be about 11\,m. For the high jump, the
athlete's center of mass would attain a maximum height of 3.5\,m. These
results are much greater than actual records. We conclude that athletes
cannot totally change their horizontal velocity into a vertical one, as
stated by the negative answers.

The negative answers bring some new questions. If the magnitude of the
velocity is assumed to be unchanged and thus the kinetic energy, why is
extra power necessary? Why does a high jumper run horizontally to jump
vertically? How do runners change their horizontal velocity into a 
velocity with a vertical component? What is conserved and what is changed at
takeoff?

\section{What limits an athlete's performance?}

There are three systems that any animal uses to produce
energy.\cite{Nelson} In the case of humans, for activities whose duration is longer 
than a few minutes, aerobic energy is produced at
a low rate by burning carbohydrates or fats. For activities that last a few
tens of seconds, the main source of energy is the breakdown of glycogen in
the absence of oxygen and the production of lactic acid. For very short
duration activities the body uses the adenosine triphosphate (ATP) stored
in muscles, and energy is supplied immediately, following the conversion of
ATP into adenosine diphosphate (ADP). 

The static force generated from muscle contraction can be very
high. However the magnitude of the force is limited by the rate that the
muscles can produce energy, which is limited by the rate that ATP is depleted
(ATP depletion is completed in a few seconds). For example, after a few
seconds sprinters no longer accelerate, but 
decelerate,\cite{MuCJP01} because they cannot
overcome the force due to air resistance. ATP is the source of energy in
all explosive sports, such as jumps, sprints, and weightlifting. Thus, if
we produce a force with a non-vanishing velocity, that is, we produce 
mechanical power, the force is limited by the rate that our muscles can
convert ATP into ADP.

To discuss the performance of a jumper, we use the results of two
observations. One is that the maximum force that athletes can produce in a
small time interval with their legs while producing work. We will take this information from 
observations of the squat. In this exercise a weight is rested on the shoulders of an athlete.
He starts from the upright position, and bends his ankles, knees,
and hips as if sitting. At the lowest position, when his thighs
are horizontal, he returns to the upright position. The maximum weight an
athlete can lift doing a squat is about 5000\,N. For a first class male athlete 
with mass 100\,kg, the maximum force that can be made with just one leg is about 3000\,N. 
(In the high and long jump, athlete's use just one leg to push their center of
mass.) This force is different from a static force.

In both the high and long jump, the height of the runner's center of mass
remains almost constant until the last stride. In the last
stride the center of mass rises about 25\,cm before the runner takes off and
his foot loses contact with the ground.

From these two observations we conclude that an 80\,kg runner
(assuming 80\,kg athlete can produce the same 3000\,N force
with each leg) can add about $(3000\--800)\,{\rm N} \times 0.25\,{\rm m} =
550$\,J to the kinetic energy. We will use this result in the
following.

An equivalent result can be obtained from an analysis of an elite 100\,m
runner. At 6\,m/s, an elite runner has an acceleration of about 5\,m/s$^2$
(see, for instance, Fig.~1 in Ref.~6). For an 80\,kg athlete this
acceleration corresponds to a mechanical power of 2400\,W. At every stride,
the runner's center of mass rises about 5\,cm. At a rate of 5\,strides/s (a
typical rate in the 100\,m dash), the up-down movement of the center of mass 
corresponds to 200\,W of additional power. Thus at each stride, the runner produces 
$(2400 + 200)\,{\rm W}
\times 0.2\,{\rm s} = 520$\,J. If we also include a small contribution due to
the air resistance, this result is consistent with the previous estimate of 550\,J. 

For female runners, the time of the 100\,m dash is about 10\% greater, and
the velocity about 10\% smaller than the male records. Thus, we estimate that
the acceleration and, hence the force, is about 20\% smaller than
that for male runners. Thus, for events that depend on velocity and force
-- as do the long and high jumps -- we expect that female performance to be
between 10\% and 20\% smaller. 

\section{Long jump}

\begin{figure}[h]
\centerline{\epsfig{figure=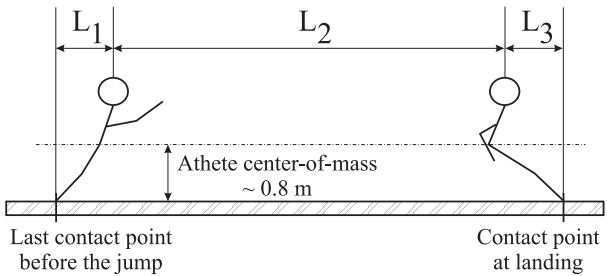,width=8cm}}
\caption[dummy0]{Schematic of the long jump. The distance of the long
jump is the sum of the takeoff ($L_1$), the flight ($L_2$), and the landing
($L_3$) distances.} \label{longjump}
\end{figure}

\begin{figure}[h]
\centerline{\epsfig{figure=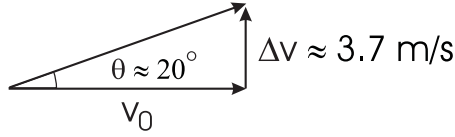,width=6cm}}
\caption[dummy0]{The long jump velocities at takeoff.}
\label{diag1}
\end{figure}

The model for the long jump is simple. After running a few seconds the
runner produces a big vertical force against the ground in the last
stride. Some details of the movement of the athlete are shown in
Fig.~\ref{longjump}. At takeoff, the runner's center of mass is about 0.4\,m
forward of the front edge of the runway, $L_1$, in Fig.~\ref{longjump}. At
landing, the runner's center of mass is behind the point where the heels hit
the sand ($L_3\simeq0.4$\,m). The difference between the vertical height of
the runner's center of mass at takeoff and at landing, which would lead to a
small difference between $L_1$ and
$L_3$,\cite{HaJB86} will be neglected. The length of the jump is
$L_1+L_2+L_3$, where $L_2$ is the flight distance. 
The horizontal velocity of the runner at the last stride, $v_0$, is about
10\,m/s. (Usually, long jumpers are good sprinters.) The best a
runner can do is to use the additional energy of 550\,J to obtain a
vertical velocity. For an 80\,kg athlete, the vertical velocity gained is
$\Delta v=3.7$\,m/s. Thus, $L_2$ is given by
\begin{equation}
L_2=\frac{2v_0\Delta v}{g}=7.6\,{\mbox m},
\end{equation}
and $L_1+L_2+L_3 = (0.4+7.6+0.4)\,{\rm m} = 8.4\,{\rm m}$. The latter result
agrees with observations between 7.4\,m and 8.8\,m 
for elite male long jumpers and 6.5\,m and 7.5\,m for elite female
jumpers.\cite{HaJB86} The takeoff angle can be calculated from the
velocity diagram in Fig.~\ref{diag1}, and we find
$\theta=\arctan(\Delta v/v_0)\sim20^\circ$. This result agrees with
observations of
$\theta\sim20^\circ\pm2^\circ$.\cite{HaJB86}

\section{High jump}

Before jumping, a high jumper runs a few meters and attains a speed of about
7.5\,m/s. Why does the jumper run horizontally before jumping vertically?
What is the mechanism used by the jumper to change her horizontal velocity
into a velocity with a vertical component? (A detailed description of the
high jump can be found in Ref.~8.)

\begin{figure}[h]
\centerline{\epsfig{figure=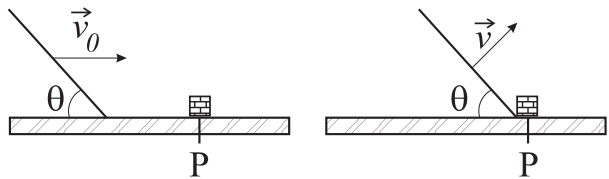,width=8cm}}
\caption[dummy0]{A bar with a horizontal velocity $v_0$ is suddenly stopped
at point P and begins to rotate, changing its initial horizontal velocity
into a new one with a vertical component (adapted from 
Fig.~2 of Ref.~8.)} 
\label{highjump}
\end{figure}

To answer these questions, consider a bar representing a jumper of length
$2\ell$, which has a horizontal velocity
$v_0$ and is inclined backward, as shown in Fig.~\ref{highjump}. At
point P, the bottom end of the bar stops suddenly, and the bar begins to
rotate around P. Due to the impact, the bar's velocity will change, but the
angular momentum with respect to P is conserved. Conservation of angular
momentum at the instant the bottom end stops gives
\begin{equation}
v=\frac34v_0\sin\theta,
\end{equation}
where $v$ is the bar's center of mass velocity immediately after the impact
and the velocity is perpendicular to the bar. Thus, the initial horizontal
velocity has changed to a velocity with a vertical component.

The horizontal, $v_h$, and vertical, $v_v$, projections of the velocity just
after the bottom end of the bar stops are
\begin{eqnarray}
\label{vh}
&&v_h=\frac34v_0\sin^2\theta,\\
&&v_v=\frac34v_0\sin\theta\cos\theta.\label{vv}
\end{eqnarray}
If the bar stops the rotation immediately after hitting P and if the
horizontal and vertical velocities at takeoff are given by Eqs.~(\ref{vh})
and (\ref{vv}), the center of mass of the bar will attain a maximum height 
(calculated by energy conservation) given by
\begin{equation}
h=\frac{1}{2g}(v_v^2+\frac2m550{\mbox J})+\ell\sin\theta.
\end{equation}
If we use $v_0$ = 7.5\,m/s and a 2\,m ($\ell = 1$\,m) bar, we obtain the
maximum
$h$ for $\theta=56^\circ$.
 
Suppose that the runner is represented by a bar moving at 7.5\,m/s
inclined backward at 56$^\circ$ (in Sec.~\ref{sec:discuss} we will show
how this motion is possible) Then the runner begins to rotate
just for a moment around point P. As in the long jump, she can add 550\,J
to the total kinetic energy. The best the higher jumper can do is to gain an
additional velocity in the vertical direction, which can
be calculated from
\begin{equation}
\frac12m(v_h^2+v_v^2)+550\,{\mbox J}=\frac12m(v_h^2+{v_v^\prime}^2),
\end{equation}
giving $v_v^\prime$ = 4.5\,m/s. The increase in her vertical velocity is 
$\Delta v=v_v^\prime-v_v=1.9$\,m/s.

\begin{figure}[h]
\centerline{\epsfig{figure=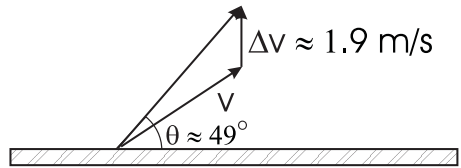,width=6cm}}
\caption[dummy0]{The high jump velocities at takeoff.}
\label{diag2}
\end{figure}

If the runner takes off with a vertical velocity equal to 4.5\,m/s, the
runner's center of mass will rise about 1.0\,m. If we add this value to the
initial position of the runner's center of mass at takeoff, which is about
1.1\,m, we obtain a total height of 2.1\,m. Because we know the horizontal
and vertical velocities, we can calculate the takeoff angle as $49^\circ$
(see Fig.~\ref{diag2}). The world record is 2.45\,m for men and 2.09\,m for
women and a typical takeoff angle is 50$^\circ$. These values suggest that
the model is reasonable.

Despite the simplification of representing a person by a bar to explain the
high jump, the pole vault is even simpler to analyze. The
pole vaulter runs about 15\,m and reaches a velocity of about 10\,m/s. Her
kinetic energy is changed to elastic potential energy of the bar, and then
the energy is changed to the vaulter's gravitational potential energy.
If we assume that the vaulter's center of mass at the beginning of the jump
is about 1.1\,m above the ground, we can use energy conservation to
calculate that the height reached by a pole vaulter is 6.2\,m. The records
are about 5 and 6\,m for men and women, respectively. This estimate is only
approximate because it does not take into account that the vaulter's center
of mass passes significantly under the bar, and it assumes a 100\%
efficiency of horizontal to vertical jump energy conversion. 

\section{Discussion and Conclusion}\label{sec:discuss}

\begin{figure}[h]
\centerline{\epsfig{figure=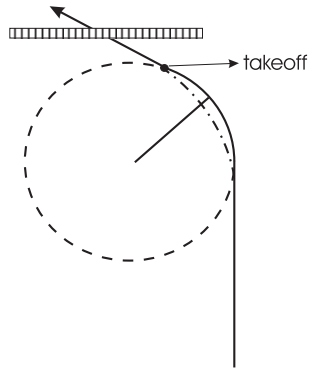,width=4cm}}
\caption[dummy0]{Trajectory developed by the athlete before high jumping. The
solid line in the circumference is the trajectory of the athlete's feet,
the dot-dashed line is the trajectory of her center of mass.} 
\label{route}
\end{figure}

Can a high jumper run with his
body inclined backward, as assumed in the model? The answer is that it
is not possible. Like everyone else, the jumper runs in a
vertical position. However, in the last strides, the high jumper makes a
circular curve with a radius of about 10\,m. Due to the centripetal force,
the jumper leans to the center of the circle and the trajectory of the
center of mass differs from the trajectory of the
footprints. These trajectories are illustrated in
Fig.~\ref{route}. At the beginning of the takeoff, corresponding to the
instant that the bottom end of the bar stops, the jumper's center of mass is
tilted toward the center of the curve, and his velocity is along the arrow
shown in Fig.~\ref{route}.\cite{DaTC95} During the next 0.2\,s, he changes
his body orientation, tilting it outside the circle, and produces a very
strong vertical force against the ground. (This model is an
oversimplification of the detailed model of Ref.~\onlinecite{DaMSSE87},
but it retains the essence of the athlete's movement at takeoff as
illustrated in Fig.~1 of Ref.~8)

In our model we assumed that the long jumper does not change her
initial velocity into a new velocity using the
same trick used by the high jumper because fixing one foot against the ground would 
cause some loss of the horizontal velocity, which is very important to the long 
jumper.

We note that the high jump differs from the vertical standing jump,
which has been studied in Ref.~\onlinecite{LiAJP01}. In the standing jump
the athlete uses both legs to jump. In the vertical standing jump the
athlete does not run before jumping and the only source of energy comes from
the leg muscles. As can be estimated from the figures of Ref.~10, the 
mechanical energy gained at takeoff by a non-athlete is about 550\,J when
using both legs; our hypothesis that an elite athlete can add 550\,J when
using only one leg seems reasonable.

\begin{acknowledgements}
This work was partially supported by the Brazilian
agency FAPESP.
\end{acknowledgements}

\end{document}